\documentclass{mem}
\usepackage{natbib}\usepackage{txfonts}\usepackage{balance}
\usepackage[dvips]{graphicx}
\usepackage{txfonts}
\usepackage[a4paper]{hyperref}
\idline{79}{3}
\begin{document}
\def\teff{$T\rm_{eff }$}
\def\kms{$\mathrm {km s}^{-1}$}

\title{The Hydrogen Ionization Front-Stellar Photosphere Interaction and the
Period-Color Relations of Variable Stars}

   \subtitle{}

\author{
S. \,Kanbur\inst{1} 
\and C. \,Ngeow\inst{2}
\and A. Nanthakumar\inst{3}
          }

  \offprints{S. Kanbur}

\institute{
Department of Physics
State University of New York at Oswego --
Oswego,
New York, NY 13126, USA
\email{kanbur@oswego.edu}
\and
Department of Astronomy,
University of Illinois,
Urbana, Illinois, IL 61801, USA
\and
Department of Mathematics,
State University of New York at Oswego --
Oswego, New York, NY 13126, USA
}

\authorrunning{Kanbur et al.}

\titlerunning{Nonlinear PL}

\abstract{
Recent evidence has emerged that the Cepheid PL relation in the LMC is nonlinear in the
sense that the existing data are more consistent with two lines of differing slope with a break at a period
of 10 days. We review the statistical evidence for this, the implications for the extra-galactic
distance scale and CMB independent estimations of Hubble's constant and briefly outline one possible
physical mechanism which could cause this nonlinearity.
\keywords{Stars: Cepheids --
Stars: atmospheres -- Stars:  -- 
Galaxy: LMC -- Cosmology: distance scale, Hubble's constant}
}
\maketitle{}

\section{Introduction}

The vast majority of work on the Cepheid based extra-galactic distance scale relies on the assumption of a linear
PL relation in the LMC in both $V$ and $I$ bands: $M_{V,I} = a + b\log P$. Recent evidence has emerged, based on a variety of
Cepheid data, that in the LMC, the Cepheid data are more consistent with two lines of differing slopes separated at a period of 10 days.
Current data indicates that this nonlinearity is present in the B,V,I,J,H bands and is marginally linear in the K. 
Here by PL relation, we mean the Cepheid PL relation at mean light.

However, we can also investigate the PL relation at different phases by phasing all light curves to a common starting epoch.
Figure 1 displays multiphase PC (left panel) and PL (right panel)
relations in $V-I$ colors and $V$ \& $I$ bands, respectively, evaluated in this way using OGLE data for LMC Cepheids.
The left panel of figure 1 also presents Galactic Cepheid PC relations.
In order to calculate these, all observations are evaluated relative to a common starting epoch.
These data have been corrected for reddening exactly as suggested by the OGLE team. The dynamic nature of both PC/PL relations
are evident as is the fact that changes in one are usually reflected in the other. At a phase of 0.82, there is a dramatic break in
both $V$ and $I$ band PL relations: perhaps the strongest evidence yet of nonlinearity in the Cepheid PL relation.
Phase 0.0 is at maximum $V$ band light. We see that at this phase the PC relation is very flat. This is similar to
Galactic PC relations at maximum light.
The mean light relation is the average of such relations over phase and so it may be expected that the behavior at phase 0.82
has some effect on the mean light PL relation. Our approach thus relies on studying PC/PL relations as a function of phase.

A nonlinearity in the form of a quadratic Cepheid PL relation has been suggested before (Marconi et al., (2005) and references therein).
However our approach is different in that the multiphase plots suggest two straight lines joined at a period of 10 days and the
physical mechanism we invoke to bring this about has not been studied before in the literature. 
\begin{figure*}
\hspace{-1.8cm}
\includegraphics[scale=0.3,height=4.5in]{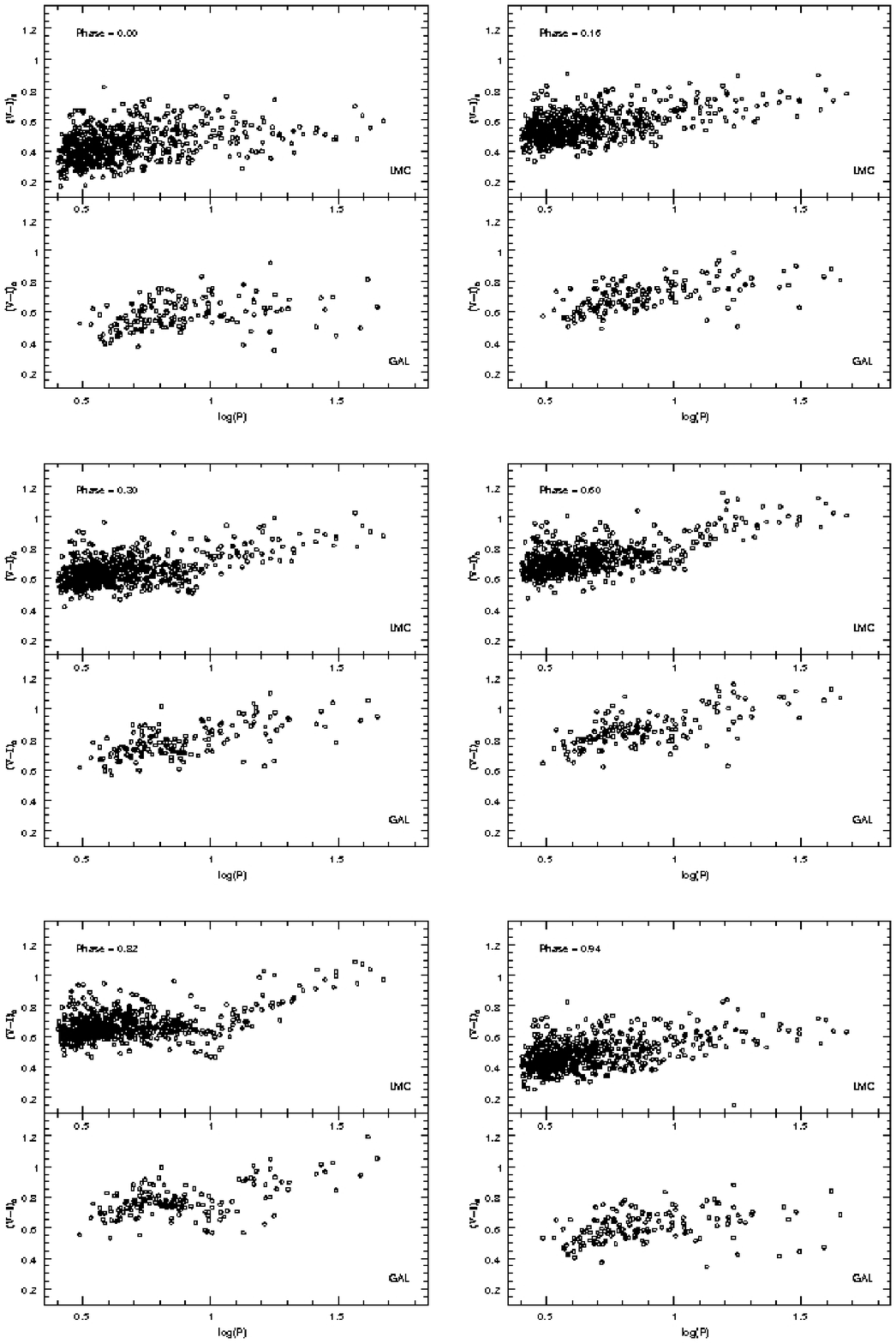}
\includegraphics[scale=0.3,height=4.5in]{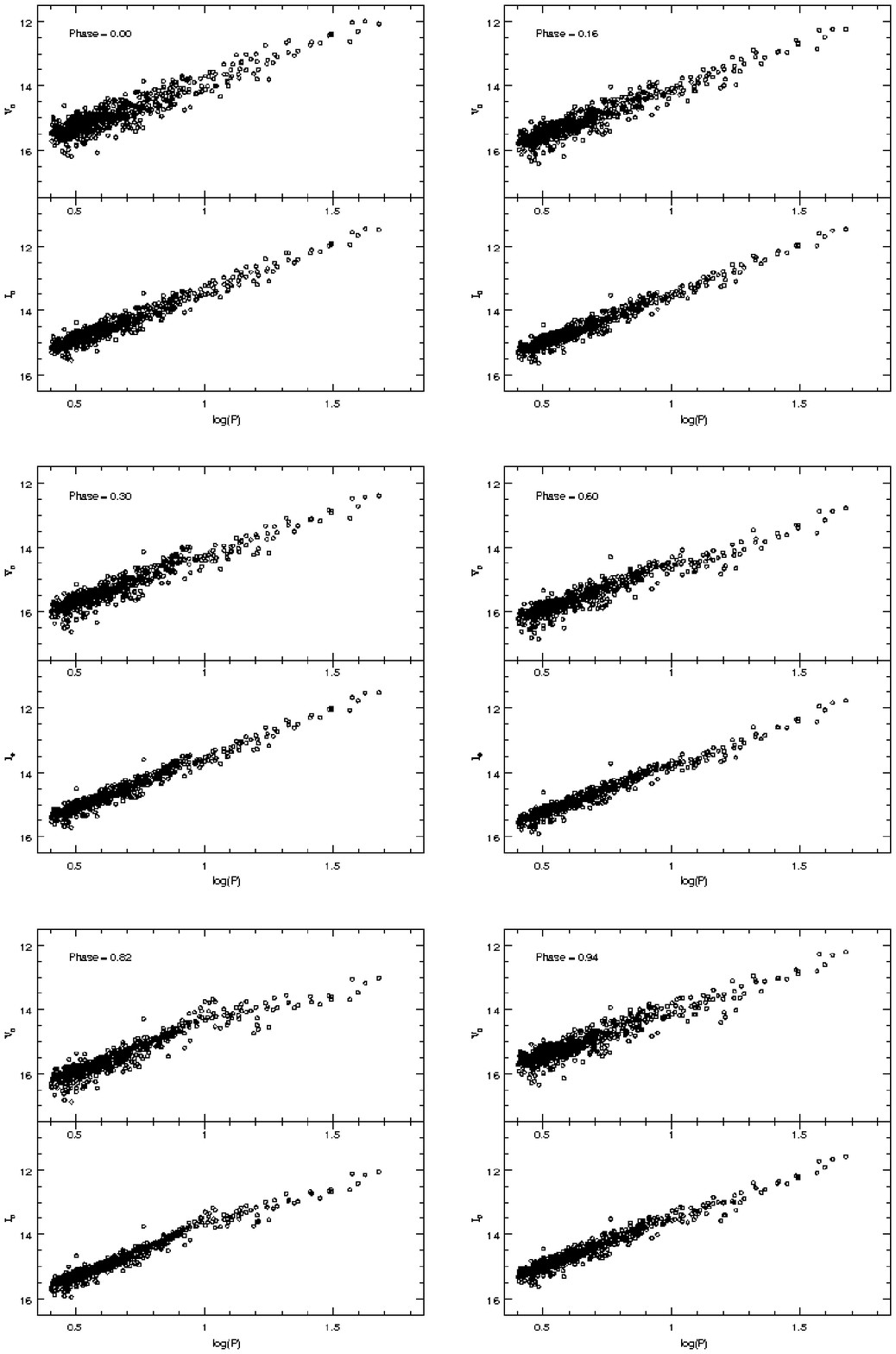}
\caption{\footnotesize
Left panel: Multiphase PC relations in $V-I$ colors for extinction corrected OGLE LMC and Galactic Cepheid data. Right panel:
Multiphase PL relations in $V,I$ for extinction corrected OGLE LMC Cepheid data. These figures are adopted from Ngeow and Kanbur (2006a).
}
\label{eta}
\end{figure*}
\begin{figure*}
\begin{center}
\includegraphics[scale=0.3,height=3.60in]{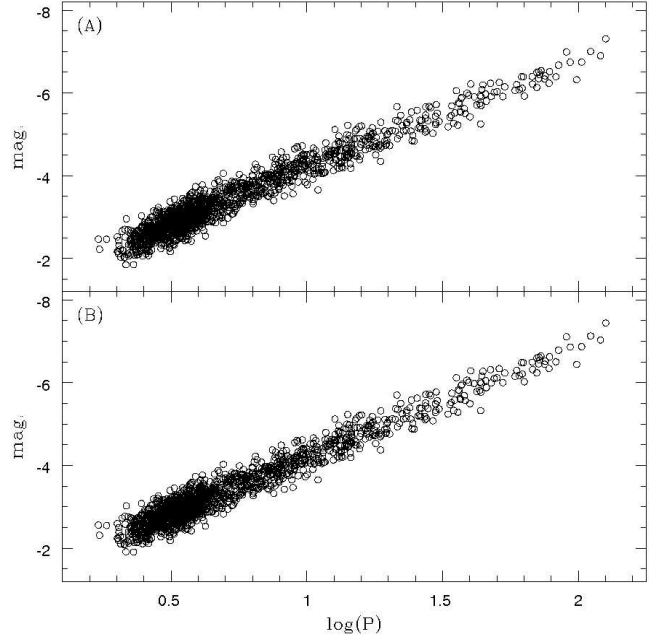}
\end{center}
\caption{\footnotesize Simulated PL relations, one of which is intrinsically linear and the other one which is intrinsically
nonlinear. They look similar and can only be distinguished by statistical tests. This figure is adopted from Ngeow and Kanbur (2006b).}
\label{eta}
\end{figure*}
\begin{figure*}
\begin{center}
{\includegraphics[scale=0.3,height=3.60in]{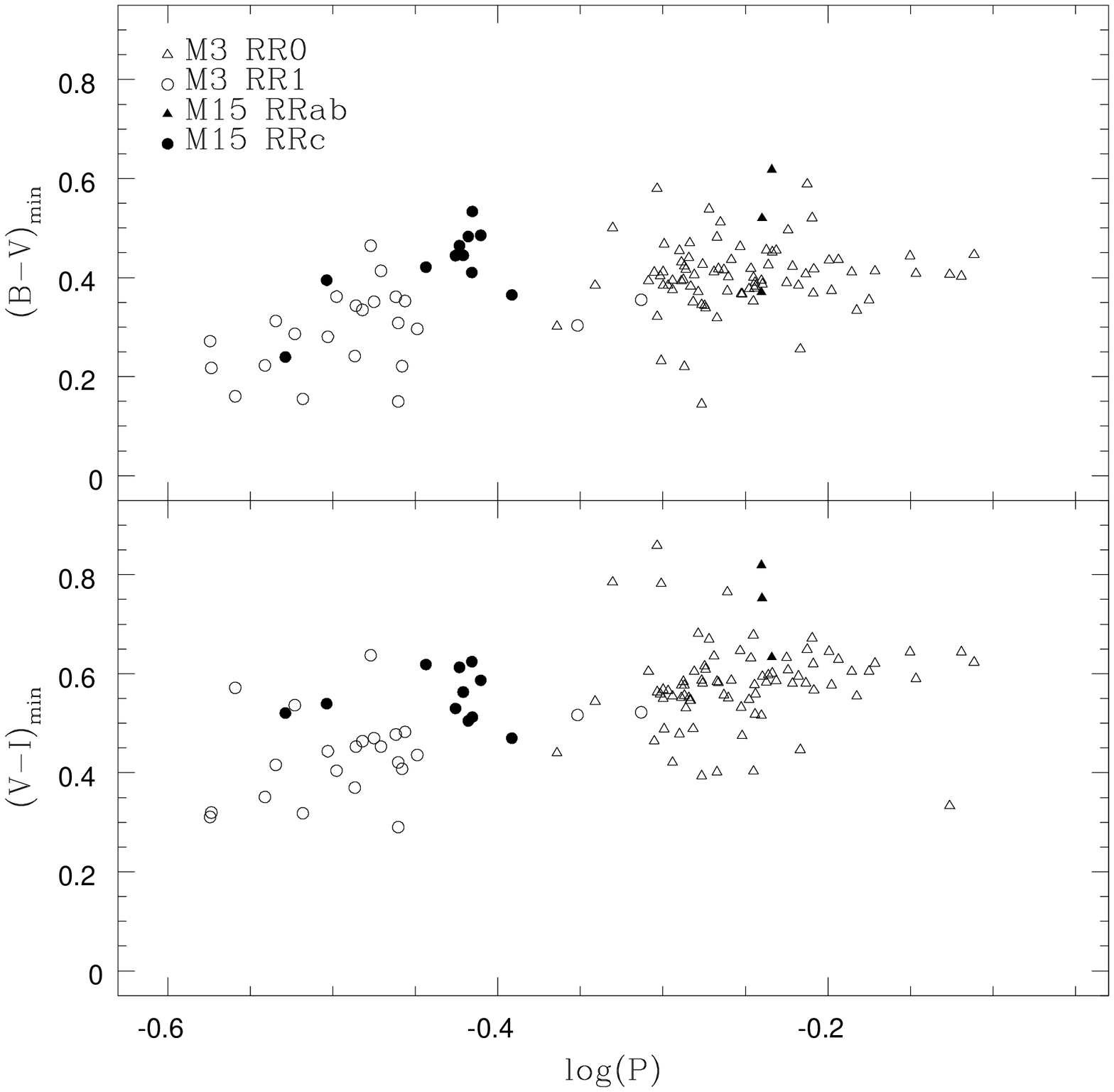}}
\end{center}
\caption{\footnotesize
RR Lyrae colors at minimum light as a function of period for M3 and M15. Overtone/fundamental pulsators are open and closed circles/triangles .}
\label{eta}
\end{figure*}
\section{Statistical tests and the nonlinearity of the PL relation}

In order to test for nonlinearity, statistical tests are required. Figure 2 displays two artificial PL
relations, with and without a break at 10 days. In both cases, the intrinsic scatter around the basic
regression line is added to mimic the real width of the instability strip. By eye it is very difficult to determine nonlinearity, yet
an $F$ test applied to these data correctly determines the intrinsically nonlinear PL relation. Further it is
insufficient to simply test for linearity/nonlinearity by simply comparing the short/long period slope and their
associated standard deviations. Estimations of slope and their standard deviations such as
"the slope is $x \pm {\delta}x$" implies the probability that
the slope is in the interval $(x-{\delta}x,x+{\delta}x)$ is $1-{\alpha}$, where $\alpha$ is the desired significance level.
If $A$ is the event that the short period slope is wrong and $B$ is the event that
the long period slope is wrong, $P(A)={\alpha}$ and $P(B)={\alpha}$. Then in comparing the long and short period slopes
purely by a comparison of the standard deviations, the probability of at least one mistake is $P(A \cup B) = P(A) + P(B) -
P(A \cap B) = 2{\alpha} - {\alpha}^2.$ If $1>{\alpha}>0$, then $2{\alpha} - {\alpha}^2 > {\alpha}$ which is the
probability the $F$ test makes a mistake.

Kanbur and Ngeow (2004), Ngeow et al (2005), Kanbur and Ngeow (2006), Kanbur et al (2007), Koen et al (2007)
applied the $F$ test, testimator, Bayesian information methods, non-parametric procedures amongst others, to the
OGLE LMC Cepheid data, MACHO LMC Cepheid data, 2MASS IR data cross-correlated with MACHO, OGLE LMC data appended with
data from Sebo et al (2002), Caldwell and Laney (1991) and Persson et al (2004) to study the linearity/nonlinearity of the
LMC Cepheid PL relation at mean light at wavelengths ranging from $B$ to $K$. All these data sets and statistical procedures
yielded the result that current data suggest the LMC Cepheid PL relation is nonlinear at a period of 10 days from $B$ to $H$ whilst the
$K$ band PL relation is marginally linear.

One possible objection to this is that there are reddening errors particularly for the longer period LMC Cepheids ie. the reddening error
is a function of period. A reddening error with period sufficient to make the mean light LMC PL relation linear in an $F$ test will result in
an extinction error as a function of period: if this same extinction error is applied to maximum light, it will imply that LMC Cepheids
get hotter/bluer at maximum light as the period increases. Such behavior is in stark contrast to Galactic Cepheids
which display a flat PC relation at maximum light. There is credible physics explaining the behavior of Cepheid PC relations
at maximum light: the interaction of the stellar photosphere and hydrogen ionization front (HIF, Kanbur Ngeow and Buchler 2004). 

Another possible objection is that there is a lack of long period data. While the number of short period Cepheids is indeed much greater than the
number of long period stars, the $F$ test takes into account both the number and nature of the data. Thus datasets with fewer numbers of
long period Cepheids determine the long period slope with a higher error which in turn makes it harder for the $F$ test to be significant.

\section{Effect on the distance scale and $H_0$}

Ngeow and Kanbur (2006c) calibrate the Hubble diagram of SNIa supernovae and estimate $H_0$ using linear and nonlinear LMC Cepheid PL
relations as calibrators. If the true relation is indeed nonlinear and a linear relation is used, an error of $1-2\%$ in $H_0$ results.
This may seem small but with significant effort being made to reduce zero-point uncertainties (Macri et al., (2006)) and important
benefits accruing from an independent Cepheid distance scale accurate to less than $5\%$, an understanding of this possible PL
nonlinearity is important. Moreover if this nonlinearity is indeed real, then it will be vital to understand its physical cause
for theories of stellar pulsation/evolution.

\section{A Possible Physical Mechanism to explain this nonlinearity}

We postulate that the same physical mechanism responsible for the flat/flatter PC relations at maximum light for Galactic and
Magellanic Cloud Cepheids affects the properties of the mean light PC relation. The HIF
and stellar photosphere move in the mass distribution as the star pulsates but they are not comoving.
The distance between them, in terms of the mass distribution, can vary according to phase, period and metallicity. If the hydrogen ionization front (HIF) and
stellar photosphere are in contact at low densities, then the temperature of the photosphere is related to the temperature at which hydrogen
ionizes. However the properties of the Saha ionization equation relevant to Cepheid envelopes means that the temperature at which hydrogen
ionizes is relatively insensitive to temperature at low densities. This can change the PC properties of the Cepheid envelope. The
amount of this change is related to period, phase and metallicity. Since changes in the PC relation affect the PL relation, we also
postulate that this mechanism can explain the nonlinear LMC PL relation: preliminary full amplitude model calculations support this idea.

\section{Extension to RR Lyraes}

The same physical mechanism can also effect the PC relation of RR Lyraes and is in fact responsible for the flat PC relation found for
RRab stars at minimum light in a number of colors. The different $L/M$ ratios and hotter effective temperatures
change the nature of the HIF-stellar photosphere somewhat but the essential idea is the same.

Figure 3 presents a plot of $(B-V)$ and $(V-I)$ color at minimum light against log period for a sample of RRab stars using data taken
by Cacciari et al (2005). Open/closed circles/triangles overtone/fundamental pulsators for M3/M15 respectively. For the fundamental mode
pulsators of M3 we clearly see a flat PC relation at minimum light. This flatness occurs due to the nature of the HIF-stellar photosphere
interaction.

\section{Conclusions}

We review recent evidence for a break in the LMC Cepheid PC/PL relation. We conclude that statistical tests are important and that current
data indicate strongly that a nonlinearity is present in the LMC Cepheid data at a period of 10 days. The multiphase relations and
consideration of PC relations at maximum light imply that it is unlikely that these nonlinearities are due to reddening errors. However
it will be important to further test a more uniform sample of LMC Cepheid data, for example, from the LMC Shallow Survey 
(Gieren 2007). We postulate that this break in the Cepheid PC and hence the PL relation
is caused by the interaction of the HIF and stellar photosphere. The variation of this interaction with period, pulsation phase
and metallicity may lead to the observed characteristics of the PL relation in the Galaxy and Magellanic Clouds.
We also note that the same physics can explain the behavior or RR Lyrae PC relations at {\it minimum light}.

\begin{acknowledgements}
SMK acknowledges support from SUNY Oswego and the American Astronomical Society through
the Chretien International research Award.
\end{acknowledgements}

\bibliographystyle{aa}

\end{document}